\documentclass[12pt. thmsa]{article}
\usepackage{amssymb}

\input{tcilatex}
\input tcilatex
\begin{document}

\author{B. Schroer and H.-W. Wiesbrock \\
Institut f\"{u}r Theoretische Physik, FU-Berlin, Arnimallee 14, Berlin,
Germany\\
schroer@physik.fu-berlin.de, wiesbroc@physik.fu-berlin.de\\
\textit{Dedicated to the memory of Harry Lehmann}}
\title{Modular Constructions of Quantum Field Theories with Interactions }
\date{November 1998}
\maketitle

\begin{abstract}
We extend the previously introduced constructive modular method to
nonperturbative QFT. In particular the relevance of the concept of ``quantum
localization'' (via intersection of algebras) versus classical locality (via
support properties of test functions) is explained in detail, the wedge
algebras are constructed rigorously and the formal aspects of double cone
algebras for d=1+1 factorizing theories are determined. The well-known
on-shell crossing symmetry of the S-Matrix and of formfactors (cyclicity
relation) in such theories is intimately related to the KMS properties of
new quantum-local PFG (one-particle \textit{polarization-free) generators}
of these wedge algebras. These generators are ``on-shell'' and their Fourier
transforms turn out to fulfill the Zamolodchikov-Faddeev algebra. As the
wedge algebras contain the crossing symmetry informations, the double cone
algebras reveal the particle content of fields. Modular theory associates
with this double cone algebra two very useful chiral conformal quantum field
theories which are the algebraic versions of the light ray algebras.
\end{abstract}

\section{Introductory Remarks}

It is known that the local field algebras of interaction-free Fermions and
Bosons can be directly obtained \cite{Schro1}\cite{Bru-Gui-Lo} from the
modular wedge-localized one-particle spaces, as a kind of inverse of the
Bisognano-Wichmann \cite{Bisognano-Wich} construction, without any reference
to the standard (semi)classical parallelism called quantization. In fact the
present paper should be viewed as one in a series of paper \cite{papers}
which attempt a new access to QFT by avoiding quantization and references to
(semi)classical approaches in the presence of interactions and using instead
modular methods. The common new finding in all these papers is an
enlargement of the symmetry concept which previously we have described in
terms of a new kind of ''hidden'' (i.e. a non-Lagrangian non-Noetherian)
symmetry structure \cite{papers}. This new concept also turns out to be
useful in the present context (section 4) which aims at an intrinsic modular
understanding of nonperturbative interactions.

In the following we will continue the investigations, which were started by
one of the authors (B.S.), on the possibility to incorporate interacting
theories into a constructive modular approach \cite{Schro1}\cite{Schro2}.
One important step in this approach was the recognition of the close
relation between the thermal Hawking-Unruh aspects of the wedge horizon on
the one hand and the subject of crossing symmetry in particle physics on the
other hand. This was subsequently (but independently) also noticed in \cite
{Nieder}. We have however some problems with the arguments given by the
latter author of such a relation; in particular we do not believe that it is
possible to give a proof of the relation without the construction of the
``polarization free generators'' of wedge algebras \cite{Schro2}. Although
these new objects were already explicitly introduced in the setting of
factorizable models in the cited previous work of one of the authors (B.
S.), we will take some pain in explaining their principle features since
their use constitutes our main new tool in the present nonperturbative
constructions. 

Our first testing ground will therefore be the d=1+1 factorizing models
which are the simplest and (apart from some less interesting low-dimensional
superrenormalizable Lagrangian models\footnote{%
In the quantization approach to interactions it is the operator short
distance dimension of the interaction density in Fockspace which separates
the models into those small superrenormalizable family (for which the
mathematical existence can be controlled), and the more interesting rest
which is not accessible by quantization methods (canonical or functional
integration). The constructive modular approach within the limit of present
knowledge does not depend in any direct way on the short distance behaviour
of particular field generators of local nets.}) presently the only
interacting properly renormalizable field theories which have a well-defined
particle (and hence a scattering) content and allow for an explicit
constructive analytic understanding. We will show that the use of PFG's goes
far beyond the understanding of the thermal KMS $\leftrightarrow $on-shell
crossing relation.

For our present purpose it is very helpful to first picture the special
position of these factorizable theories within the set of general d=1+1
Wightman field theories as representing massive superselection classes in
the following way. Imagine that we remove all particles by spatially
separating their (centre of) wave packets ``on shell'' i.e. in the
scattering matrix. Using as a hindsight the analytic singularity structure
of the multiparticle creation (annihilation) threshold, as well as the
relation between momentum space singularities and fall-off behavior in its
Fourier transform, we are led to believe that these inelastic multiparticle
matrix elements possess decreasing properties in this extreme cluster limit
relative to the elastic ones. Even more, the direct multiparticle elastic
processes in this limit is expected to become asymptotically negligible as
compared with the two particle elastic contribution \cite{Schro2}. In this
way we obtain factorizing S-matrices which fulfill the Yang-Baxter
consistency equations without reference to quantization and infinitely many
conserved Noether currents; with other words the Y-B structure is not
imposed but rather derived from the general principles of QFT in the above
explained extreme cluster limit. Although, as a result of lack of detailed
knowledge about the analytic structure of admissable S-matrices of QFT this
picture is mathematically not rigorous, it is very useful for the motivation
behind our investigation and for an intrinsic local quantum physical
understanding of the position of factorizable models within general QFT
which removes some of their ``freak appearance''.

In fact it is one of the folklore theorems (but in this case rigorously
provable under suitable analytic assumptions) that for d=1+3 theories the
limiting S-matrix is S=1. The limiting field theories are then expected to
be free fields which only record the superselection charges (internal
symmetries) of the original theory. However in d=1+1 the simplification does
not go all the way to free theories since the cluster property is not
capable to separate the elastic two-particle T-matrix from the identity in
the decomposition S=1+T (intuitively because two particles always meet in
x-space or mathematically because the two particle energy momentum delta
function in d=1+1 happens to be identical to the two particle inner product
delta function). All the multiparticle elastic scattering then takes place
through two-particle scattering and the Yang-Baxter relation follows as a
physical consistency requirement (and is not a mathematical imposition) and
allows together with the unitarity and crossing symmetry (closely linked to
the TCP symmetry) to separate and solve the construction of the S-matrix
from the field theory. This is a peculiar feature of the so constructed
simple (but nontrivial) ''factorizing representative'' of d=1+1; in any
other field theory including nonfactorizing d=1+1 models, there is no
possibility to extract a pure S-matrix bootstrap setup from the off-shell
field theory; rather the S-matrix must be determined together with the
fields or (in our setting) the net of local algebras.

The principle purpose of this paper, which generalizes the previous findings
of one of the present authors \cite{Schro2} concerning the modular
understanding of crossing symmetry, is a more detailed explanation of the
useful constructive role of so called ``PFG'' (one-particle \textbf{p}%
\textit{olarization-}\textbf{f}\textit{ree} wedge \textbf{g}\textit{enerators%
}) i.e. on-shell but nevertheless wedge localized operators $F(x),$ which
create one-particle states without additional pair contributions\footnote{%
We have called such operators on previous occasions ``vacuum-polarization
free'' and used the abbreviation FWG in \cite{Schro2}, but a better
terminology used in the following is to call them ``one-particle-\textit{%
polarization free generators}'' (PFG). Wheras the notion of
vacuum-polarization which originated with conserved currents in free field
theory in the work of Heisenberg and Weisskopf, the present concept of
``one-particle polarization'', i.e. the impossibility of creating pure one
particle states (without pair admixtures) from local operators is totally
characteristic for interacting theories (and may be used as a good intrinsic
definition of the latter).} known from the famous vacuum polarization in
QFT. In fact they constitute a system of generators for the wedge algebra.
These ``fields'', unlike pointlike local fields (either associated with
Lagrangians or with the Wightman framework), have a \textit{quantum} (but no
classical) \textit{localization} i.e. they remain wedge localized even if
smeared with test functions which have sharper localized supports inside the
wedge region. With quantum localization, this sharper localization has to be
obtained from the \textit{intersection of wedge algebras}; the intersected
algebras in turn have their own new generators. The PFG's are more
noncommutative than the pointlike localized fields and they are in fact not
derivable from any known quantization approach. But on the other hand they
are essential in our nonperturbative construction which uses (quantum)
modular wedge localization; although they are themselves nonlocal, they have
a very precise relation to local fields. In some vague sense one may say
that the semilocal intermediary PFW's have a milder short distance behavior
than pointlike local fields, but their use is in agreement with physical
principles, whereas the use of cutoff fields is not and, therefore the
explanation in terms cutoffs is somewhat misleading. In any case the
existence of the theory in the modular approach becomes detached from the
``threatening'' short distance behavior of pointlike field coordinates. The
borderline between renormalizabe/nonrenormalizable theories looses its
meaning and instead the existence of local theories becomes linked with the 
\textit{nontriviality of intersections of certain algebras}.

For the aforementioned factorizing models one can argue that their mass
shell Fourier transforms obey the Zamolodchikov-Faddeev algebra \cite{Schro2}%
. In fact the construction of PFG's in factorizing models may be done in a
spirit similar to the perturbative construction of local fields in the
retarded (Yang-Feldman) formalism. The PFG's viewed in this way are on-shell
analogues of the off-shell interacting fields, but unlike the incoming free
fields they carry information about the interaction. Whereas the Z-P
algebraic structure is essential for the present analytic control of the
bootstrap-formfactor- approach, the concept of \textit{quantum wedge
localization} relates to a vastly more general algebraic structure of the
commutator of $F$ with its modular reflected $JFJ$ which is controlled by a
crossing symmetric (a symmetry around the imaginary rapidity $\frac{1}{2}%
i\pi )$ operator $M(\theta )$ as will be explained in the third section. The
understanding of the concept of quantum wedge localization for these objects
is the main objective in section 3.. Different from the usual locality of
local fields or nets of algebras which exists independently of the
particular representation, the new quantum locality depends crucially on the
presence of the vacuum representation and one-particle states (more
precisely on the standard assumptions which led to the validity of rigorous
time-dependent scattering theory). It is only after the modular construction
of a local net has been accomplished, that one may change the vacuum
representation with another locally normal one.

After the role of the PFG's for quantum wedge localization has been
presented, we describe the formal aspects of their intersections which leads
to double cone algebras in section 4. Via suitably defined relative
commutants we construct two light ray theories whose conformal rotation is a
``hidden'' symmetry. The problem of nontriviality as well as the the
explicit construction of the double cone algebra is greatly facilitated in
terms of these ``smaller'' light ray algebras.

\section{An Intuitive Argument for Existence of PFG's}

In this section we will present some intuitive (mathematically nonrigorous)
arguments why we expect one-particle polarization free wedge generators
(PFG's) to exist, although we are presently only able to rigorously show
this for d=1+1 factorizing systems.

Let us briefly recollect how the perturbative approach deals with
interactions. The formulation which is most appropriate for the present
purpose is that of Stueckelberg, Bogoliubov and Shirkov \cite{Bog} and
partially also that of Weinberg \cite{Wein}; here we do not need the more
sophisticated version of Epstein and Glaser \cite{E-G}. The first step from
Wigner particles to Fockspace and free fields and the second step of
implementing interactions by forming invariant Wick-ordered local
polynomials of composite free fields and defining transition operators 
\begin{equation}
S(g,h)=T\exp i\int \left\{ g(x)W(x)+h(x)A(x)\right\} dx  \label{1}
\end{equation}
where for simplicity we used a symbolic notation and wrote the polynomial
interaction density as a monomial $W$ and denotes by $h$ the source
function(s) of the basic free field coordinate(s) $A$ in terms of which we
specify the Fock space and the $W.$ As usual T denotes time-ordering. These
operator functionals of the testfunctions in Fock space fulfill the so
called Bogoliubov axiomatic and it is of no relevance to us whether this
axiomatics has nonperturbative solutions or not; the reader is entitled to
take the most pessimistic view concerning their existence. Formally these
time-ordered exponentials would represent the scattering operator $S_{sc}$
if the test functions approach the constant function on Minkowski space. If
only this ``on-shell'' value of this time-ordered operator would be known,
there would be no possibility of computing off-shell local fields. Therefore
Bogoliubov et al. \cite{Bog} assume (either by functional dependence or in
some other way) that there exists an off-shell extrapolation of $S_{sc}$
which can be related (by their functional derivative formalism which is not
part of scattering theory) to local fields and is given by (\ref{1}). This
formalism then leads to an expression for the outgoing free field and to the
Yang-Feldman equation 
\begin{eqnarray}
A_{out}(x) &=&A_{in}(x)+\int \Delta (x-x^{\prime })j(x^{\prime })dx^{\prime
}\equiv S_{sc}^{*}A_{in}(x)S_{sc} \\
A(x) &=&A_{in}(x)+\int \Delta _{ret}(x-x^{\prime })j(x^{\prime })dx^{\prime }
\nonumber \\
j(x) &=&K_{x}S^{*}(h)\frac{\delta }{\delta h(x)}S(h)\mid _{h=0}  \nonumber
\end{eqnarray}
Here $\Delta $ is the mass-shell Pauli-Jordan commutator function, whereas $%
\Delta _{ret}$ is the off-shell retarded function which is formally the
on-shell commutator function multiplied with the step function in time. The
alternative right hand way of writing in the first line indicates that the
on shell restriction of the $\int \Delta (x-x^{\prime })j(x^{\prime })$
contribution is determined by the on-shell object $%
S_{sc}^{*}[A_{in}(x)S_{sc}],$ i.e. the definition via the off-shell $S(g,h)$
is not needed. The crucial question is now if these formulas can be used in
a suggestive manner for the construction of a semilocal wedge localized $%
F(x) $ which like $A_{in}(x)$ is still on-shell, but also like $A(x)$ it
carries informations about interactions. Assume for simplicity that we are
in the situation of a factorizable model with a diagonal S-matrix. Such an
elastic S-matrix in terms of the analytic phase shift $\delta ,$ with $%
\delta (\theta )\varepsilon (\theta )=\delta _{phys}(\left| \theta \right|
), $ would be represented by an exponential of a quadrilinear term in the
incoming field 
\begin{equation}
S_{sc}=\exp i\frac{1}{2}\int \int \delta (\theta -\theta ^{\prime
})\varepsilon (\theta -\theta ^{\prime }):\rho (\theta )\rho (\theta
^{\prime }):d\theta d\theta ^{\prime }
\end{equation}
where the $\rho (\theta )$ is the rapidity space charge density (resp.
particle number density in our case of selfconjugate particles). From this
we read off the relation between $S_{sc}$and $a_{out}$%
\begin{eqnarray}
&&a_{out}(\theta )=S_{sc}^{*}a_{in}(\theta )S_{sc}= \\
&=&a_{in}(\theta )\exp i\int_{-\infty }^{\infty }\delta (\theta -\theta
^{\prime })\rho (\theta ^{\prime })d\theta ^{\prime }  \nonumber
\end{eqnarray}
It is now suggestive to try the following nonlocal field as a candidate for
a wedge localized PFG field \cite{Schro2} 
\begin{eqnarray}
F(x) &=&\frac{1}{\sqrt{2\pi }}\int \left\{ Z(\theta )e^{-ipx}+Z^{*}(\theta
)e^{ipx}\right\} d\theta  \label{F} \\
Z(\theta ) &=&a_{in}(\theta )\exp i\int_{-\infty }^{\theta }\delta (\theta
-\theta ^{\prime })\rho (\theta ^{\prime })d\theta ^{\prime }  \nonumber
\end{eqnarray}
The analogue of the off shell split of $\Delta $ into $\Delta _{ret}$ and $%
\Delta _{av}$ is the split of the on shell rapidity space integrals of the
S-matrix. This analogy gains more credibility from the remark that the
exponent is indeed some generalization of an integral over the current $j(x)$
a fact which is particularily evident \cite{Schro2} in the case of the
Federbush- and Thirring- model There are also indications that this analogy
between ``off-shell space time and on-shell velocity'' or rather its
rapidity logarithm may be helpful in understanding certain ``rapidity space
cluster properties'' \cite{Kou-Mu} which are expected to select a subclass
of local pointlike operators which recently were observed in formfactors of
certain models. And last not least, we expect this analogy to be helpful in
a future unraveling of the structure of PFG's in the nonperturbative
analysis of non-factorizing QFT's.

Clearly the positive and negative frequency parts of the above $F(x)$ obey a
Zamolodchikov-Faddeev algebra \cite{Zam} (see (\ref{Z}) of the next
section). Now the generalization to factorizing systems with more general
nondiagonal S-matrices is obvious: generalize the structure on the algebraic
side and prove that the so obtained PFG fields are indeed wedge localized in
the sense of quantum localization. The motivation for our notation should be
obvious; whereas the symbol $F$ denotes the more general PFG operators which
are expected to exist in each QFT, its positive and negative frequency
components are denoted by the symbol $Z^{\#}$ since for d=1+1 factorizing
theories these generalized ``creation and annihilation'' operators fulfil
the Zamolodchikov-Faddeev algebra. Our previous intuitive argument
restricted to the context of factorizing models relates the new concept of
PFG with the Z-F algebra in a nontrivial and useful way.

In order to achieve a more detailed understanding of this connection, we
need to develop some more formalism. We first return briefly to the wedge
localization of free fields.

\section{Generators of Quantum Wedge Localization}

In order to verify the modular wedge localization for the $Z_{W}$ operators,
we first look back at the free n- particle wedge localization \cite{Schro2}
and introduce some additional notation. Using free fields the wedge
localization spaces may be written: 
\begin{eqnarray}
\left| \psi ^{(n)}\right\rangle &=&\int \hat{\psi}%
^{(n)}(x_{n},x_{n-1},...x_{1}):A(x_{1})...A(x_{n}):\Omega \\
supp\hat{\psi}^{(n)} &\subseteq &W^{\otimes n},\,\,A(x)=\frac{1}{\sqrt{2\pi }%
}\int e^{-ipx}a(\theta )+e^{ipx}a^{*}(\theta )  \nonumber
\end{eqnarray}
where $W$ denotes the right hand wedge and the operator involves a Wick
product of free fields $A$ $($for the sake of notation chosen scalar and
selfconjugate)$.$ Instead of the vectors $p$ and $x$ we use their rapidity
parametrization: 
\begin{equation}
p=m\left( 
\begin{array}{l}
\cosh \theta \\ 
\sinh \theta
\end{array}
\right) ,\,\,x=\rho \left( 
\begin{array}{l}
\sinh \chi \\ 
\cosh \chi
\end{array}
\right) \,for\,\,x\in W
\end{equation}
We also prefer the more intrinsic way of writing which avoids the use of
field coordinates (which are not unique) and uses the momentum space
creation and annihilation operators which are directly linked to the Wigner
representation theory of irreducible particle representations: 
\begin{eqnarray}
\left| \psi ^{(n)}\right\rangle &=&A_{\psi ^{(n)}}\Omega  \label{loc} \\
A_{\psi ^{(n)}} &=&\int_{C}...\int_{C}\psi ^{(n)}(\theta _{n},\theta
_{n-1},...\theta _{1}):a(\theta _{1})...a(\theta _{n}):d\theta
_{1}...d\theta _{n}  \nonumber
\end{eqnarray}
The used path notation $C$ is a self-explanatory notation which generalizes
the rapidity representation of the wedge localized fields \cite{Lash}: 
\begin{eqnarray}
A(\hat{f}) &=&\int f(\theta )a(\theta )+\int f(\theta -i\pi )a(\theta -i\pi
)\equiv \int_{C}f(\theta )a(\theta )  \label{wedge} \\
&&a(\theta -i\pi )\equiv a^{*}(\theta ),\,\,f(\theta -i\pi )=\bar{f}(\theta )
\nonumber
\end{eqnarray}
$C$ consists of the real $\theta -$axis and the parallel path shifted down
by $-i\pi $ and it is only the function $f$ which is analytic in the strip $%
-\pi <Im\theta <0$ and conjugate-symmetric (i.e. fulfilling the Schwarz
reflection principle) around the $\func{Im}\theta =\frac{1}{2}i\pi $ line.
For the operators this is only a notational convention and implies no
analyticity\footnote{%
Operators in QFT, either in x-space or in momentum space, are never
analytic, although some unfortunate notation and terminology especially in
chiral conformal QFT suggests this. For more remarks see \cite{Schro2}.}.
The $f$ analyticity is equivalent to the localization property of $\hat{f},$
and the analytic properties of the state vectors $\psi $ and the path
notation in (\ref{loc}) is a n-variable generalization of (\ref{wedge}) In
the application to the vacuum in (\ref{loc}) of course only the creation
contribution from the lower rim of the strip survives. It is easy to see
that these vectors fulfill the modular localization equation \cite{Schro2}: 
\begin{eqnarray}
&S\left| \psi ^{(n)}\right\rangle =\left| \psi ^{(n)}\right\rangle &
\label{equ} \\
&S=J\Delta ^{\frac{1}{2}},\,\,J=TCP,\Delta \,^{it}=U(\Lambda (2\pi t))\,& 
\nonumber
\end{eqnarray}
The antilinear unbounded Tomita involution $S$ (be aware to avoid confusions
with the notation for the S-matrix!) consists (for Bosons) of the TCP
reflection $J$ and the analytically continued (by functional calculus) wedge
affiliated Lorentz boost $U(\Lambda (\chi ))$ and the analytical strip
properties guarantee that the localized vectors $\left| \psi
^{(n)}\right\rangle $ are in the domain of $\Delta ^{\frac{1}{2}}$ and hence
of $S.$ The action of $\Delta ^{\frac{1}{2}}$ corresponds to the
continuation to the lower rim and the action of $J$ is just complex
conjugation in momentum space (for selfconjugate situations): 
\begin{equation}
\psi ^{(n)}(\theta _{n},\theta _{n-1},...\theta _{1})\stackrel{S}{%
\rightarrow }\overline{\psi ^{(n)}(\theta _{n}-i\pi ,\theta _{n-1}-i\pi
,...\theta _{1}-i\pi )}
\end{equation}
so that modular localization equation (\ref{equ}) states that the value of
the wave function on the $-i\pi $-shifted boundary equals the complex
conjugate of the upper boundary. Note that the hermiticity of $A_{\psi
^{(n)}}$ implies the reality condition $\overline{\psi ^{(n)}(\theta
_{n},\theta _{n-1},...\theta _{1})}=\psi ^{(n)}(\theta _{1},...\theta _{n})$
without analytic properties. For Bosons the wave functions are of course
symmetric. In case of Fermions it is well known that the Tomita reflection $%
J $ has a Klein twist in addition to the $TCP.$

The closure $H_{R}^{(n)}$of the real subspace of solutions of (\ref{equ})
contains the spatial part of modular wedge localization. By applying the
generators of the wedge algebra (\ref{wedge}) to the vacuum n-times, we
generate a dense set of localized state vectors in the complex space which
turns out to be the n-particle component of the well known Reeh-Schlieder
set of vectors. This dense space becomes the Hilbert space $%
H_{R}^{(n)}+iH_{R}^{(n)}$ if one forms the closure in the graph norm: 
\begin{equation}
\left\langle \psi _{2},\psi _{1}\right\rangle \equiv \left( \psi _{2},\psi
_{1}\right) +\left( S\psi _{1},S\psi _{2}\right)
\end{equation}
Let us now pass to the case with interaction. For simplicity of notation we
assume that the S-matrix of the factorizing interacting model describes the
interaction of only one kind of particle (neutral, without bound states.
Different from the free case, it follows from scattering theory \cite{Schro2}
that $J$ carries all the interaction whereas the $\Delta ^{it}$ remains
unchanged. Whenever necessary we will add a suffix $0$ for the free
(incoming) objects. With this notation we have: 
\begin{equation}
J=S_{sc}J_{0},\,\,\Delta ^{it}=\Delta _{0}^{it}
\end{equation}
where $S_{sc}$ denotes the S-matrix. The realization that the scattering
connection between asymptotic (e.g. incoming) free fields and interacting
fields (resp. algebras) keeps the unitary representations of the connected
part of the Poincar\'{e} group unmodified and only changes those
disconnected components which contain the anti-unitary time reversal is
well-known \cite{St-Wi}. What is new is the realization that the
interpretation within the modular framework attributes to the S-matrix the
property of a relative modular invariant for wedges. To see this, one only
has to remember that thanks to the the Bisognano-Wichmann connection of the
Lorentz boost and the TCP operation with modular theory, the relation $%
J=S_{sc}J_{0}$ is just the TCP transformation law of the S-matrix with the
TCP operator being expressed in terms of the Tomita conjugation. In d=1+1
the $J$ is identical with the TCP operator whereas in higher dimensions it
is different by a $\pi $-rotation which commutes with the $S_{sc}.$

In analogy with the free n-particle Hilbert spaces $H_{R}^{(0)}$ we make the
Ansatz:

\begin{eqnarray}
\left| \psi ^{(n)}\right\rangle &=&A_{\psi ^{(n)}}\Omega \\
A_{\psi ^{(n)}} &=&\int_{C}\psi ^{(n)}(\theta _{n},\theta _{n-1},...\theta
_{1}):Z(\theta _{1})...Z(\theta _{n}):  \nonumber
\end{eqnarray}
Here the $Z^{\prime }s$ are also mass shell annihilation and creation
operators but with more complicated nonlocal commutation relations: 
\begin{eqnarray}
Z^{\#}(\theta _{1})Z^{\#}(\theta _{2}) &=&S_{sc}(\theta _{1}-\theta
_{2})Z^{\#}(\theta _{2})Z^{\#}(\theta _{1}),\,\,Z^{\#}\equiv Z\,\,or\,\,Z^{*}
\label{Z} \\
Z(\theta _{1})Z^{*}(\theta _{2}) &=&S_{sc}^{-1}(\theta _{1}-\theta
_{2})Z^{*}(\theta _{2})Z(\theta _{1})+\delta (\theta _{1}-\theta _{2}) 
\nonumber \\
crossing &:&S_{sc}(\theta )=S_{sc}(i\pi -\theta )=S_{sc}^{*}(\theta -i\pi ) 
\nonumber
\end{eqnarray}
In a different and more formal context this algebraic structure (the ''Z-F''
algebra) was introduced by Zamolodchikov and later completed (by adding the $%
\delta -$function term) by Faddeev. The states they generate are connected
with the in-states by 
\begin{eqnarray}
&&Z^{*}(\theta _{1})...Z^{*}(\theta _{n})\Omega =a^{*}(\theta
_{1})...a^{*}(\theta _{n})\Omega \\
&&for\,\,\theta _{1}>\theta _{2}>...>\theta _{n}  \nonumber
\end{eqnarray}
and the identification for permuted orders following from the Z-F algebra.
It should be stressed that these consistent identifications are only
tentative, pending on their reproduction by (LSZ, Haag-Ruelle)
time-dependent scattering theory applied to the local operators which are
still under construction. In the case of nondiagonal S-matrices it has not
been possible to guess an operator formula on the basis of the above
identification of states; more comments on this problem can be found in the
next section. If we treat the $Z^{\#}(\theta )$ in analogy to mass shell
creation and annihilation operators, we (following our intuitive discussion
in the previous section) should form the ``field'': 
\begin{eqnarray}
F(x) &=&\frac{1}{\sqrt{2\pi }}\int \left( e^{-ipx}Z(\theta
)+e^{ipx}Z^{*}(\theta )\right) d\theta \\
Z^{*}(\theta ) &=&Z(\theta -i\pi )
\end{eqnarray}
It is of course no surprise that this field is noncausal. It is however a
bit better than that; it is ''wedge local'' i.e. the smeared operator $%
F(f)=\int F(x)f(x)d^{2}x$ with supp$f\in W$ generates a $^{*}$-algebra of
the interacting theory localized in the wedge. In formula 
\begin{equation}
\left[ JF(f)J,F(g)\right] =0,\,\,\,\,\text{supp}f,g\in W  \label{w.1.}
\end{equation}
To prove this one first notices that the $Z(\theta )^{\#}$ commutes with the 
$JZ(\theta )^{\#}J$ underneath the Wick-ordering$.$ The reason is that the
exponentials involve integrals over number density which extend over
complementary rapidity regions. The numerical phase factors which originate
from the commutation of these exponential factors with the $a^{\#\prime }s$
mutually compensate. There remains the contraction between the
pre-exponential $a^{\#\prime }s$ which leads to 
\begin{equation}
\int_{C}\bar{f}(\theta )g(\theta -i\pi )\exp i\int \delta _{sc}(\theta
-\theta ^{\prime })n(\theta ^{\prime })d\theta ^{\prime }
\end{equation}
Shifting the integration by -$i\pi $ as required by the lower boundary of $%
C, $ and using the crossing symmetry (particle-antiparticle Schwartz
reflection symmetry around -$i\frac{\pi }{2})$ in the form: $\delta
_{sc}(\theta -i\pi )=\delta _{sc}(-\theta )+2\pi ni,$ we see that this
contraction is equal to that in the opposite order (which has the negative
exponential). Again one easily verifies that this argument goes through if
one only uses the commutation and vacuum annihilation properties of the $%
Z^{\#\prime }s.$

The mathematical control over the $Z^{\#}(\theta )$ and $F(f)$-operators is
not more difficult than that of the standard creation and annihilation
operators $a^{\#}(\theta )$ and the smeared free fields $A(f).$ Since the
vacuum is annihilated by the $Z^{\prime }s,$ the n-particle vectors are
generated by n-fold application of $\int Z^{*}(\theta )f(\theta )d\theta $
onto $\Omega .$ With $S^{(2)}$ being a crossing symmetric solution of the
bootstrp program, the n-particle component of the action of this operator on
a state vector $\psi $ is then 
\begin{eqnarray}
&&(\int Z^{*}(\theta )f(\theta )d\theta \psi )^{(n)}(\theta _{1},...\theta
_{n}) \\
&=&\frac{1}{\sqrt{n}}\sum_{i=1}^{n}f(\theta
_{i})\prod_{k=1}^{i-1}S^{(2)}(\theta _{k}-\theta _{i})\psi ^{(n-1)}(\theta
_{1},..\check{\theta}_{i},..\theta _{n})  \nonumber
\end{eqnarray}
where we used $\check{\theta}$ for the omission of a variable and where we
have suppressed the indices on the wave functions on which the two-body
S-matrix acts. As in the bosonic case, the norm of this wave function obeys
the standard inequality involving the number operator $\mathbf{N}$%
\begin{equation}
\left| \left| \int Z^{*}(\theta )f(\theta )d\theta \psi \right| \right|
\leqslant \left| \left| f\right| \right| \left| \left| \mathbf{N}\psi
\right| \right|
\end{equation}
and hence the closability of the $Z^{\#\prime }s$ and the selfadjointness of 
$F(f)$ for real test functions follows in a well-known manner \cite{Bo-Zi},
despite the fact that the total particle space 
\begin{equation}
\mathcal{H}=\oplus _{n}\mathcal{H}_{n}  \label{space}
\end{equation}
is not manifestly identical to a Boson/Fermion Fock space. In fact for none
of the calculations we need a formula for $Z^{\#}$ in terms of free incoming
fields as in (\ref{F}), the general algebraic relations (\ref{Z}) together
with the vacuum annihilation property is all we use for the construction of
wedge localized generators in the space (\ref{space}). The check of the
Tomita relation 
\begin{equation}
J\Delta ^{\frac{1}{2}}A\Omega =A^{*}\Omega
\end{equation}
can be done directly on products of the generators which are associated to
the von Neumann algebras or alternatively by checking the KMS relation (see (%
\ref{KMS})) together with the transformation into the commutant by $J.$

Contrary to the previous free case, the sharpening of the support of the
test function does not improve the localization within the wedge. This is
equivalent to the statement that the reflection with $J$ does not create an
operator which is localized at the geometrically mirrored region in the
opposite wedge $W^{\prime }.$ It only fulfills the commutation relation with
respect to the full $W^{\prime }.$ In fact the breakdown of parity
covariance is important for the existence of such nonlocal but
wedge-localized fields since fields which are covariant under all
transformations are expected to be either point local or completely
delocalized (i.e. not even in a wedge). This breakdown of parity covariance
for the Z-fields does not mean that the theory violates parity symmetry but
only that these auxiliary fields only fulfill that symmetry and localization
requirement which are expressible in terms of wedge algebras without
reference to possible pointlike local field generators. We will later see
that any sharper localization requires the operator to be an \textit{%
infinite power series} in the $Z^{\prime }s:$%
\begin{equation}
A=\sum \frac{1}{n!}\int_{C}...\int_{C}a_{n}(\theta _{n},\theta
_{n-1},....\theta _{1}):Z(\theta _{1})....Z(\theta _{n}):  \label{series}
\end{equation}
where we again used the previously explained path notation. The sharper
localization leads in fact to relations between the $a_{n}^{\prime }s$
(later) which, with one coefficient being different from zero forces higher
ones to be nonvanishing as well$.$ Therefore the PFG's only serve as a
natural basis for smaller than wedge algebras, they themselves are not
generators of these algebras. These statements comply with the physical idea
that whereas the noncompact wedge region is big enough to allow the
identification of particle states (it contains Lorentz boosts as
automorphisms), it is on the other hand small enough to contain no
annihilators as required by having a unique relation between vector states
and operators (the Reeh-Schlieder relation) which is the prerequisite for
the modular theory. This delicale balance is broken if one passes to compact
localized algebras which favor the field side with the sharp particle number
(and iven the one-particle aspect) loosing its observable meaning.

The proof of (\ref{w.1.}) used the crossing symmetry of the S-matrix which
appears in the Z-F algebra (\ref{Z}). Note that since the commutations of $%
Z^{\prime }s$ produce unitary factors with the same two-particle S-matrix,
the $a_{n}$ must compensate this unitary factors upon commuting $\theta
^{\prime }s.$ 
\begin{equation}
a_{n}(\theta _{1},...\theta _{i},\theta _{i+1},..\theta
_{n})=S_{sc}^{(2)}(\theta _{i}-\theta _{i+1})a_{n}(\theta _{1},...\theta
_{i+1},\theta _{i},..\theta _{n})  \label{c r}
\end{equation}
In this respect the phase factors are like statistics terms. However since
the $a_{n}$ have analytic properties in the multi-$\theta $ strip (actually
for compact localization the analytic region is much bigger), these phase
factors must be consistent with the univaluedness in the analytic domain.

We now show that the KMS property relative to the boost is equivalent to the
crossing symmetry of the $S^{(2)}$-coefficients in the Z-F algebra. The
first nontrivial correlation function which deviates from that of free
fields is the 4-point function. The KMS property reads 
\begin{equation}
\left\langle F(f_{1^{^{\prime }}})F(f_{2^{^{\prime
}}})F(f_{2})F(f_{1})\right\rangle _{therm}=\left\langle F(f_{2^{^{\prime
}}})F(f_{2})F(f_{1})F(f_{1^{^{\prime }}}^{2\pi i})\right\rangle _{therm}
\label{KMS}
\end{equation}
where the superscript $2\pi i$ is the imaginary KMS shift in the boost
parameter. Each side is the sum of two terms, the direct term associated
with 
\begin{eqnarray}
F(f_{2})F(f_{1})\Omega &=&\int f_{2}(\theta _{2}-i\pi )f_{1}(\theta
_{1}-i\pi )Z^{*}(\theta _{1})Z^{*}(\theta _{2})\Omega +c-number\cdot \Omega 
\nonumber \\
&=&\int f_{2}(\theta _{2}-i\pi )f_{1}(\theta _{1}-i\pi )S^{(2)}(\theta
_{2}-\theta _{1})a^{*}(\theta _{1})a^{*}(\theta _{2})\Omega +c\Omega
\end{eqnarray}
and the analogous formula for the bra-vector. For the inner product there
are two contraction terms consisting of direct and crossed contraction (in
indices 1 or 2) of the $a^{\#}s.$ Only the second one gives an S-matrix
factor in the integrand. The c-number term an the left hand side cancels the
direct term on the right hand side. The equality of the crossed terms on
both sides gives (using the denseness of the analytic wave functions) 
\begin{equation}
S^{(2)}(\theta _{2}-\theta _{1})=S^{(2)}(\theta _{1}-\theta _{1^{\prime
}}+i\pi )\mid _{\theta _{1^{\prime }}=\theta _{2}}
\end{equation}

i.e. one obtains the above crossing relation for the two particle S-matrix.
Higher inner products involve products of S-matrices and it is easy to see
that the KMS condition for the Z-F algebra is equivalent to the crossing
property of the S-matrix. The presence of operators $A$ which are localized
in a e.g. double cone (without loss of generality within a wedge) does not
influence the validity of the KMS condition. 
\begin{eqnarray}
&&\left\langle F(f_{1^{^{\prime }}})F(f_{2^{\prime }})...F(f_{m^{^{\prime
}}})AF(f_{n})...F(f_{2})F(f_{1})\right\rangle _{therm}= \\
&=&\left\langle F(f_{2^{\prime
}})..F(f_{m})AF(f_{n})..F(f_{2})F(f_{1})F(f_{1^{^{\prime }}}^{2\pi
i})\right\rangle _{therm}  \nonumber
\end{eqnarray}

The rapidity space formulation of this KMS condition will turn out (see next
section) to be the cyclicity relation for the formfactor of that local
operator which hitherto \cite{Karowski} was a special consequence for
factorizing systems of the formally derived crossing symmetry which in turn
follows from the LSZ scattering theory together with analytic assumptions.

It is not difficult to find an axiomatic generalization of PFW operators
from this illustrative model theory to the general situation of d=1+1 wedge
localized generators. Let us assume that we have a PFG operator $F(f)$ and
its J-transform $JF(g)J$ such that the commutator in rapidity space after
splitting off the $f,g$ factors which we denote by $M$ commutes with the
Poincar\'{e}-group. This assumption is certainly satisfied in the
factorizing case (see below). Then the fluctuation terms do not contribute
to the commutator and we obtain for the structure of the matrix elements of
the $F-JFJ$ commutator between multiparticle states (the $F-JFJ$
formfactors) 
\begin{eqnarray}
&&^{out}\left\langle p_{1}^{\prime }...p_{n}^{\prime }\right| \left[
JF(f)J,F(g)\right] \left| p_{1}...p_{m}\right\rangle ^{in}  \label{F-JFJ} \\
&=&\int \bar{f}(\theta )g(\theta -i\pi )M(\theta _{1}^{\prime },...,\theta
_{n}^{\prime };\theta -i\pi ;\theta _{1}...,\theta _{m})d\theta  \nonumber \\
&&-\int \bar{g}(\theta )\bar{f}(\theta +i\pi )M^{J}(\theta _{1}^{\prime
},...,\theta _{n}^{\prime },\theta ;\theta _{1}...,\theta _{m})d\theta 
\nonumber
\end{eqnarray}
Here we assumed the validity of the following commutation relations 
\begin{eqnarray}
\left[ Z^{J}(p),Z(p^{\prime })\right] &=&0=\left[ Z^{J*}(p),Z^{*}(p^{\prime
})\right]  \nonumber \\
\left[ Z^{J}(p),Z^{*}(p^{\prime })\right] &=&2\omega \delta (p-p)M(p)
\label{cr}
\end{eqnarray}
i.e. we assume that the relative commutators between $Z^{\prime }s$ and $%
Z^{J\prime }s$ generalize those between free field annihilation/creation
operators in momentum space in that they are Poincar\'{e}-invariant
operators (but not necessarily multiples of the identity as for free
fields). In terms of the F's we used the following notation 
\begin{eqnarray}
\left[ JF^{(-)}(f)J,F^{(+)}(g)\right] &=&\int \int \bar{f}(\theta ^{\prime
})g(\theta -i\pi )\delta (\theta ^{\prime }-\theta )M(\theta -i\pi )d\theta
^{\prime }d\theta \\
M(\theta _{1}^{\prime },...,\theta _{n}^{\prime };\theta -i\pi ,\theta
_{1}...,\theta _{m}) &=&^{out}\left\langle \theta _{1}^{\prime },...,\theta
_{n}^{\prime }\right| M(\theta -i\pi )\left| \theta _{1}...,\theta
_{m}\right\rangle ^{in}  \nonumber
\end{eqnarray}
Note that the second contribution on the right hand side of (\ref{F-JFJ})
results from the identity 
\begin{eqnarray}
&&^{out}\left\langle \theta _{1}^{\prime },...,\theta _{n}^{\prime }\right|
\left[ F^{(-)}(g),JF^{(+)}(f)J\right] \left| \theta _{1}...,\theta
_{m}\right\rangle ^{in} \\
&=&^{out}\left\langle \theta _{1}^{\prime },...,\theta _{n}^{\prime }\right|
JJ\left[ F^{(-)}(g),JF^{(+)}(f)J\right] \left| \theta _{1}...,\theta
_{m}\right\rangle ^{in}  \nonumber \\
&=&\overline{^{out}\left\langle \theta _{1}^{\prime },...,\theta
_{n}^{\prime }\right| J\left[ JF^{(-)}(g)J,F^{(+)}(f)\right] J\left| \theta
_{1}...,\theta _{m}\right\rangle ^{in}}  \nonumber
\end{eqnarray}

If now $M$ is ''crossing symmetric'' around $\func{Im}\theta =-\frac{1}{2}%
i\pi $ as the left hand side in (\ref{cr}) suggests, i.e. $M$ and $M^{J}$
are strip-analytic and 
\begin{eqnarray}
&&M(\theta _{1}^{\prime },...,\theta _{n}^{\prime };-\frac{1}{2}i\pi
-\vartheta ;\theta _{1}...,\theta _{m}) \\
&=&M^{J}(\theta _{1}^{\prime },...,\theta _{n}^{\prime };-\frac{1}{2}i\pi
+i\vartheta ;\theta _{1},...,\theta _{m})  \nonumber
\end{eqnarray}
then we have established wedge locality i.e. the $F$ are really wedge
generators and hence the validity of the following theorem

\begin{theorem}
The quantum wedge localization of the PFG $F$ is equivalent to the existence
and crossing symmetry of their $Z_{W}$-$JZ_{W}J$ formfactors.
\end{theorem}

There remains the task to \textit{prove this crossing symmetry for the
factorizing models}. From the structure of the Z-F algebra and the
definition one concludes that $M(\theta )$ has the form 
\begin{equation}
M(\theta )=T(\theta )
\end{equation}
where $T(\theta )$ is Lorentz ($\theta $-translation) covariant, leaves the
vacuum unchanged and fulfills the following commutation relation with the
Z's 
\begin{equation}
T(\theta )Z(\theta ^{\prime })=S^{(2)}(\theta -\theta ^{\prime })Z(\theta
^{\prime })T(\theta )
\end{equation}
which together with $T(\theta )\Omega =\Omega $ fix the operator. In the
case of a diagonal S-matrix the T has the following expression in terms of
the incoming fields 
\begin{equation}
T(\theta )=e^{i\int_{-\infty }^{+\infty }\delta (\theta -\theta ^{\prime
})n(\theta ^{\prime })d\theta ^{\prime }}
\end{equation}
Here as before we left it to the reader to convince himself that one can do
all the computations without knowing the formulas for Z's or T's in terms of
incoming fields, which covers the general factorizable setting with many
particles which may form multiplets. We will later on comment on how to
obtain such an operator relation for this case.

\section{The Construction of the Double-Cone Algebra}

Finally we to indicate (again for the simplest one particle model) how one
proceeds from the quantum localized Z's to the double cone algebras.
Formally we have to look for solutions of 
\begin{equation}
JBJ=A,\,\,\,A,B\in \mathcal{A}(\mathcal{O}_{a})  \label{dc}
\end{equation}
the equation for the elements of the algebraic intersection 
\begin{equation}
\mathcal{A}(\mathcal{O}_{a})=U(-\frac{a}{2})\mathcal{A}(W)U(\frac{a}{2})\cap
U(\frac{a}{2})\mathcal{A}(W^{\prime })U(-\frac{a}{2})
\end{equation}

The wedge algebra is the von Neumann algebra associated with the $^{*}$%
-algebra 
\begin{equation}
\mathcal{A}_{formal}(W)=\left\{ A\mid A=\sum_{n}\frac{1}{n!}\int_{C}d\theta
_{1}...\int_{C}d\theta _{n}a_{n}(\theta _{n},\theta _{n-1}....\theta
_{1}):Z(\theta _{1})....Z(\theta _{n}):\right\}  \label{formal}
\end{equation}

where the $a_{n}$ are meromorphic in the lower strip. Here we call a
would-be von Neumann operator algebra $\mathcal{A}_{formal}$ if we only
study aspects of sesquilinear forms i.e. modulo convergence properties
required by bona fide operators. These coefficient functions are related to
the KMS property of the previously introduced mixed correlation functions 
\begin{eqnarray}
&&\left\langle AF(f_{1})...F(f_{n})\right\rangle _{therm} \\
A &\in &\mathcal{A}(O_{a}),\,\,\,\,F(f_{i})\in \mathcal{A}(W_{a})  \nonumber
\end{eqnarray}
Originally the KMS property gives the analyticity in the regions of ordered
imaginary parts for monomial vectors $\prod_{i}Z(f_{i})\Omega ,$ but as a
result of the denseness of the analytic $f_{i}^{\prime }s,$ one retains
meromorphy (analyticity modulo poles) for the coefficient functions of
general n$^{th}$ components $a_{n}$ of wedge localized operators. In fact
the expansion of $A$ in terms of $F^{\prime }s$ may be directly written in
x-space 
\begin{eqnarray}
A &=&\sum \frac{1}{n!}\int ...\int \hat{a}%
(x_{1},...,x_{n}):F(x_{n})...F(x_{1}):  \label{rap} \\
&=&\sum \frac{1}{n!}\int_{C}...\int_{C}a(\theta _{1},...\theta
_{n}):Z(\theta _{n})...Z(\theta _{1}):  \nonumber \\
&&\left\langle AF(f_{1})...F(f_{n})\right\rangle _{therm}=\int ...\int \hat{a%
}_{n}(x_{1},...,x_{n})f(x_{1})...f(x_{n})+lower\,\,terms  \nonumber
\end{eqnarray}
The compact localization allows to extend the meromorphy to the product of
the $\theta $-planes. As before in the case of the KMS property of the pure
F correlation functions was equivalent to the crossing symmetry of the
S-matrix, the KMS property of the mixed functions involving one $A$ with
compact localization (or a field $A(x)$ with $x\in W)$ to a crossing
symmetry relation, the so-called cyclicity relation in the rapidity
variables $\theta .$

\begin{theorem}
The KMS property of the mixed wedge correlation functions is equivalent to
the cyclicity relation of the coefficient functions in the rapidity
representation 
\begin{equation}
a_{n}(\theta _{2},...,\theta _{n},\theta _{1}-2\pi i)=a_{n}(\theta
_{1},...,\theta _{n})
\end{equation}
\end{theorem}

The proof uses the rapidity representation (\ref{rap}). For the highest
coefficient in the KMS relation one obtains the cyclicity relation by
straightforward computation again (as in the previous S-matrix case) using
the density of the boundary values of the strip analytic test functions $%
f_{i}(\theta _{i})$

For many considerations it is safer to argue directly in terms of the
thermal correlators. in x-space than with the momentum rapidity $%
a_{n}^{\prime }s.$ Even the following derivation of the pole structure
should be transcribed into the thermal correlators, but we will present it
in the more familiar momentum space rapidity form. In order to avoid
questions about the necessity to introduce ``fused'' $Z(\theta )^{\prime }s$
and $F^{\prime }s$ for fused (bound states\footnote{%
The notion of bound states carries connotations of Q.M. which are somewhat
misleading in relativistic field theories with (apart from free fields) a
virtual particle polarization structure. The correct hierarchy is that of
basic versus fused charges whose low-lying carries are (infra)particles.})
particles into the expansions (\ref{formal}) and other questions related to
the completeness of $Z^{\prime }s,$ we will simply assume the absence of
such particles and defer a more general discussion to a future publication.

Now $B$ $\in \mathcal{A}_{formal}(W)^{\prime }$ has the analogue
representation to (\ref{formal}) in terms of $Z^{J}$ and coefficient
functions $b_{n}$ which are upper strip-meromorphic. Shifting the apex of
the opposite wedge by a translation a into $W$ means that we multiply the $%
b_{n}$-coefficients with $\exp i\sum p_{i}(\theta _{i})a.$ Let us call the
shifted b's $b_{n}^{(a)}.$ The formal double cone algebra $\mathcal{A}%
_{formal}(\mathcal{O}_{a})$ is defined as $\mathcal{A}_{formal}\cap U(a)%
\mathcal{A}_{formal}(W)^{\prime }U^{*}(a).$ In order to compute matrix
elements of $Z^{J}$ within vector states created by $Z^{\prime }s$ we use
the cumulant formula 
\begin{equation}
\exp i\int \sum_{i}\delta (\theta _{i}-\theta ^{\prime }\dot{)}a^{*}(\theta
^{\prime })a(\theta ^{\prime })d\theta ^{\prime }=:\exp \int (1-\Pi
_{i}S(\theta _{i}-\theta ^{\prime }))a^{*}(\theta ^{\prime })a(\theta
^{\prime })d\theta ^{\prime }:
\end{equation}
The computation is facilitated by using the following equivalent modular
characterization of the intersection algebra in terms of either $A\in 
\mathcal{A}(W_{\pm a})$ and 
\begin{eqnarray}
A &=&A_{s}+iA_{s}  \nonumber \\
JA_{s}J &=&A_{s}
\end{eqnarray}
where we decomposed a general operator of the double cone algebra into a $J$%
-selfconjugate and antiselfconjugate part. Here $A\,$belongs to the shifted
algebra, but $J$ is the Tomita involution of the original wedge algebra,
i.e. is our previous $J.$ Since the $J$ does not mix even and odd terms, we
will assume that $A$ has only even terms in its power series.

Matching first the diagonal matrixelements for the $A_{s}^{\prime }s$ and $%
JAJ^{\prime }s,$ we obtain the following recursion for the boundary values
of the meromorphic functions: 
\begin{eqnarray}
&&a_{0}=\bar{a}_{0} \\
&&a_{2}(\theta _{1},\theta _{2}-i\pi )=\bar{a}_{2}(\theta _{1},\theta
_{2}+i\pi )+\delta (\theta _{1}-\theta _{2})(1-S(\theta _{1}-\theta
_{2}-i\pi ))a_{0}  \nonumber \\
&&a_{4}(\theta _{1},\theta _{2},\theta _{3}-i\pi ,\theta _{4}-i\pi )=\bar{a}%
_{4}(\theta _{1},\theta _{2},\theta _{3}+i\pi ,\theta _{4}+i\pi )  \nonumber
\\
\,\,\, &&+\delta (\theta _{1}-\theta _{3})(1-S(\theta _{1}-\theta _{4}-i\pi
)S(\theta _{2}-\theta _{3}+i\pi )S(\theta _{1}-\theta _{2}))a_{2}(\theta
_{2},\theta _{4}-i\pi )  \nonumber \\
&&+3\,\,more\,\,such\,\,terms  \nonumber \\
&&a_{6}(\theta _{1},......)\,\,\,\,\,=\,\,\,etc.  \nonumber
\end{eqnarray}
with an analogous recursion for the odd coefficient functions. Taking into
account the meromorphic properties of the coefficient functions, these $%
\delta $-function terms in the boundary value mean that the meromorphic
functions have poles if two rapidities coalesce modulo $i\pi .$ The matching
conditions also contain the possibility of continued extension into the
tensor product of complex planes. In this way one obtains meromorphic
functions in the multi-$\theta $ plane which have a Paley-Wiener type of
increase in the imaginary direction which is related to the size $a$ of the
double cone. Again one easily notices that these consideration can be done
solely on the basis of the algebraic properties of the $Z^{\#}$ operators
together with their annihilation property of the vacuum vector. Possible
poles in the coefficient functions from other fused particle states appear
in crossing symmetric pairs and compensate; only the above kinematical poles
enter the formal determination of the double cone algebras $\mathcal{A}(%
\mathcal{O}_{a}).$ Hence our method applies to the general case of
factorizable theories with admissable nondiagonal bootstrap S-matrices \cite
{Kar}.

In addition to the above kinematical poles there are poles inside the
physical strip associated with ``bound states'' \footnote{%
Strictly speaking the hierarchy of elementary versus bound states is limited
to quantum mechanics and looses its meaning in the presence of (virtual)
vacuum polarization. The only hierarchy which remains meaningful is that
between basic and fused charges; on the level of particles QFT practices
``Nuclear Democracy'' i.e. interactions couple all channels which are not
separated by superselection rules.}. As in the case of wedge localization
they do not contribute to the structure of $\mathcal{A}_{formal}(\mathcal{O}%
_{a})$ as a result of their pairwise occurrence due to crossing symmetry.

As a result, we obtain the well-known ``kinematical pole structure'' of
formfactors (the coefficients in the series (\ref{series}) known from the
work of Smirnov \cite{Smirnov}\cite{Lash} (based on recipes) and later
abstracted from LSZ scattering formalism in \cite{Karowski}. In our approach
this structure (as the crossing symmetry) is equivalent to the $J$%
-invariance of the symmetrically placed double cone algebra.

\begin{theorem}
Necessary and sufficient for the series (\ref{series}) to fulfill (\ref{dc})
is that the coefficient functions are meromorphic in the multi $\theta $%
-plane with certain fall-off behavior in imaginary direction and following
pole structure on the boundary of the old strip 
\begin{eqnarray}
&&a_{n+2}(\theta +i\pi +i\varepsilon ,\theta ,\theta _{1},...,\theta
_{n})\mid _{\varepsilon \rightarrow 0}  \label{resid} \\
&=&\frac{1}{\varepsilon }\left[ 1-\prod_{i=1}^{n}S^{(2)}(\theta -\theta
_{i})\right] a_{n}(\theta _{1},...,\theta _{n})  \nonumber
\end{eqnarray}
Since each pair of $\theta ^{\prime }s$ can always be transported into this
canonical position by the application of (\ref{c r}), there is no loss of
generality in this way of writing.
\end{theorem}

The formfactors of pointlike fields at the origin which have appeared in the
literature \cite{Smirnov} \cite{Karowski} result by taking the limit a$%
\rightarrow 0,$ prescribing the Lorentz transformation property of the would
be field and defining a convention which picks a particular (composite)
field in the remaining infinite dimensional field space. Since fields at the
origin are sesquilinear forms and not operators, they are improper members
of $\mathcal{A}_{formal}(\mathcal{O}).$ The operator problems show up in the
computation of correlation functions of these fields. In fact there are two
formidable tasks in the Smirnov-Karowski-Weisz approach. On the one hand one
must construct a basis in field space (the analogue to the Wick-polynomials
in the free field case) which even in the simplest nontrivial so-called
SinhGordon model led to presently intractable looking problems about
symmetric polynomials in the $x_{i}\equiv \exp \theta _{i}$ \cite{Kou-Mu}.
But this is not enough, according to Wightman one also has to take care of
the operator problems of the smeared fields who'sss prerequisite is the
existence of the correlation functions. Apart from those special cases with
a constant S-matrix as the Ising- and Federbush models, this has not been
achieved. In the latter cases one notices immediately that there are
quadratic expressions in the $Z^{\prime }s$ for which the iteration yields a
vanishing factor from the factor in front of $a_{2}$ on the right hand side
in (\ref{resid}). Since the operator aspects of such quadratic expressions
as the energy-momentum tensor are easily controlled, it is clear that not
only $\mathcal{A}_{formal}(\mathcal{O})$ but also $\mathcal{A}(\mathcal{O})$
is nonempty. For rapidity dependent S-matrices the iteration cannot stop and
the construction of even a single operator causes serious mathematical
problems.

Using the algebraic approach, one only faces a ``light'' version of these
problems. The reason is that the modular construction liberates the
existence proof of a model from the technical burden of field coordinates
and sometimes even insures the existence of local copies of symmetry
operators. One only has to show that the double cone algebras are not
trivial i.e. contain operators which are not multiples of unity. This step
would then allow to omit the subscript ``formal'' in the above space of
sesquilinear forms and write $A\in \mathcal{A}(\mathcal{O}_{a})=U(-\frac{a}{2%
})\mathcal{A}(W)U(\frac{a}{2})\cap U(\frac{a}{2})\mathcal{A}(W^{\prime })U(-%
\frac{a}{2})$. Let us first note that a trivial double cone algebra ($%
\mathcal{A}(\mathcal{O}_{a})=C\cdot \mathbf{1})$ would violate several
properties of Local Quantum Physics. On the one hand the additivity property 
\begin{eqnarray}
W &=&\cup _{d}\mathcal{O}(0,d) \\
\mathcal{A}(W) &=&\vee \mathcal{A}(\mathcal{O}(0,d))  \nonumber
\end{eqnarray}
where $\mathcal{O}(0,d)$ is a family of double cones inside $W$ which have a
common left corner with $W,$ does not hold. Furthermore the so called split
property cannot hold in such a case for it is one of the consequences \cite
{Haag} that a double cone algebra cannot be just a multiple of the identity
because it must contain at least a local version of the symmetries of the
theory e. g. an algebraic analog of the energy momentum tensor. But in our
situation we need an argument which refers only to the commuting wedge
algebras $\mathcal{A}(W),\mathcal{A}(W^{\prime })\equiv J\mathcal{A}(W)J$
and the d=1+1 Poincar\'{e} group. If we were to assume the split property
for two dimensional wedges \cite{Mueger}, we would immediately obtain the
unitary equivalence of the double cone algebra to a tensor product of wedge
algebras 
\begin{equation}
\mathcal{A}(\mathcal{O}_{a})\simeq \mathcal{A}(W_{\frac{a}{2}})\otimes 
\mathcal{A}(W_{-\frac{a}{2}}^{\prime })
\end{equation}
so that the nontriviality of the double cone algebra would be manifest. This
elegant argument has however one drawback; all factorizable models are very
different from superrenormalizable models which are locally normal with
respect to free theories and therefore inherit the split property of the
latter and as a result we have doubts about the validity of the split
property for wedges for factorizing models. Therefore we will try yet
another argument based on the idea that the double cone algebra possesses a
nontrivial ``light cone reduction''. The mathematical formulation of this
idea starts from the relative commutant associated with the following
inclusion 
\begin{eqnarray}
&&\mathcal{A}(W_{a_{\pm }})\subset \mathcal{A}(W) \\
&&i.e.\,\,\mathcal{A}(W_{a_{\pm }})^{\prime }\cap \mathcal{A}(W)
\end{eqnarray}
Here $W$ is the d=1+1 standard wedge and $W_{a_{\pm }}$ are the upper and
lower wedges obtained by having $W$ slide into itself along the upper/lower
light ray ($\pm $horizon) with the distance $a_{\pm }$. The geometric
position of the associated relative commutant $\mathcal{A}(W_{a_{\pm
}})^{\prime }\cap \mathcal{A}(W)$ is the 1-dimensional interval on the
upper/lower light ray which starts at the origin and ends at $a_{\pm }.$ It
may be considered to be the limiting case of the 2-dim. relative commutant
which is obtained by shifting the $W$ into itself so that the upper/lower
horizons of the shifted wedge do not yet touch the standard wedge. From
these upper/lower relative commutants we may form 
\begin{equation}
\mathcal{M}_{\pm }\equiv \bigcup_{t}\Delta _{W}^{it}\left( \mathcal{A}%
(W_{a_{\pm }})^{\prime }\cap \mathcal{A}(W)\right) \subset \mathcal{M}\equiv 
\mathcal{A}(W)
\end{equation}
This strict inclusion reflects the interpretation as being the conformal
light ray limit. A manifest way of exposing this structural property of this
limiting theory on the light ray is to emphasize that the improvement of
symmetry on the light ray\footnote{%
The fact that this net on the light ray which is indexed by intervals is not
only covariant under light ray translations and scale transformations but
also under conformal rotations follows from \cite{Wies}.} \cite{Sew} to full
conformal symmetry goes together with a decrease in the size of the
cyclically generated space from the vacuum 
\begin{eqnarray}
\mathcal{H}_{\pm } &\equiv &\mathcal{M}_{\pm }\Omega \subset M\Omega \equiv 
\mathcal{H=}P_{\pm }\Omega \\
E_{\pm }(\mathcal{M}) &=&P_{\pm }\mathcal{M}P_{\pm }=\mathcal{M}_{\pm } 
\nonumber
\end{eqnarray}
The projector $P_{\pm }$ onto this smaller space is associated with a
conditional expectation $E_{\pm },$ in agreement with the fact that the two
theories share the same modular group. To recover the full algebra from its
light cone pieces, we have to combine $\mathcal{M}_{\pm }.$ There are two
ways of doing this 
\begin{eqnarray}
\mathcal{M}_{(+,-)} &=&\mathcal{M}_{+}\vee \mathcal{M}_{-}  \label{double} \\
\mathcal{M}_{\otimes } &=&\mathcal{M}_{+}\otimes \mathcal{M}_{-}  \nonumber
\end{eqnarray}
The second way belongs to the construction of two-dimensional conformal QFT
from their chiral components whereas the first one gives the equality of the
combined light cones to reproduce the original massive algebra in case that
the vacuum is cyclic since $clos(\mathcal{M}_{(+,-)}\Omega )=clos(\mathcal{M}%
\Omega )$ $\leftrightarrow $ $\mathcal{M}_{(+,-)}=\mathcal{M}.$

In order to see what is going on in the case of factorizing models in more
detail, let us write down the commutation relation which characterize those
operators $A\in \mathcal{M}$ (\ref{series}) which commute with (the
generators of) $\mathcal{M}_{+}$ 
\begin{equation}
\left[ A,\int f_{a_{+}^{\prime }}(x)Z(x)d^{2}x\right] =0,\,\,a_{+}^{\prime
}\geq a_{+}
\end{equation}
Clearly the n$^{th}$ order contribution in the $Z(\theta )$ power series
receives two contributions, one with a coefficient function of the form 
\begin{equation}
\int_{C}d\theta (\Pi S(\theta _{i}-\theta )-1)a_{n-1}(\theta _{1},..\theta
_{n-1})e^{imu\exp \theta }
\end{equation}
which results from the product of S-matrices from commuting the $Z(\theta )$
to the left hand side, as well as contraction terms 
\begin{equation}
\int_{C}d\theta \int_{C}d\theta _{n}a_{n+1}(\theta _{1},...\theta
_{n-1},\theta _{n},\theta )e^{imu\exp \theta }
\end{equation}
A pause of thought reveals that this relative commutator condition does not
tell us anything on the pole structure of the coefficient functions which we
do not know already from theorem (\ref{resid}), however due to the presence
of only one light cone coordinate the vanishing of the commutator is
identical to problems in the algebraic formulation of chiral nets on the
line with tranlation and scale covariance. The one dimensional operator
version of the Paley-Wiener theorem gives a much better control on the
remaining meromorphy properties. The reduction to the light ray fields gives
a powerful tool for the transition from the wedge algebras to the compactly
localized double cone algebras. We will give a detailed account of its
application to factorizable models. Here we will be content to present some
more structural properties in the sequel.

The light ray reductions are bona fide conformal nets since the
translation-dilation situation together with the modular inclusion $\mathcal{%
M}_{a_{\pm }}^{\prime }\subset \mathcal{M}$ is known to correspond to a
SL(2,R)-covariant local conformal net \cite{Guido}. It is very important not
to equate this net with the in a naive way neither with the (divergent)
light cone restriction nor with the (Lagrangian) light cone quantization%
\footnote{%
The formal light cone quantization procedure leads to to a kind of quantum
mechanics as a result of the surpression of vacuum polarization \cite{Heinz}%
. In none of the presentations of the light cone formalism an attempt is
made to reconstruct the original local fields from the light ray data. We
believe that this is not possible without the algebraic reprocessing by
modular methods as we used in this paper.} of fields. For factorizing
theories whose local generating fields are more singular for short distances
than free fields (they are not superrenormalizable as it would be required
by local normality), the canonical structure is not defined and the light
cone restriction is divergent since the field near to the + light cone does
generally not become independent on the - light cone variable. It is only
through the associated algebraic net (generated by the local fields) that
the light ray restriction is defined. This algebraic reprocessing is even
necessary in superrenormalizable theories since the light cone quantization
in the presence of interactions is nonlocal (it is a PFG) even when it is
well defined. Only by reprocessing the nonlocal data via relative commutants
one has a chance to reconstitute a local net on the light ray. With other
words the modular method is essential for understanding and physically
interpreting light ray physics. Simple examples show that the light ray
restriction generates a space which is genuinly smaller than the total
Hilbert space.

The simplest illustration of this mechanism is obtained by looking at the
free massive Fermion field 
\begin{eqnarray}
\left\{ \psi (x),\psi ^{*}(y)\right\} &=&\left( \gamma _{\mu }\partial ^{\mu
}-m\right) i\Delta (x-y) \\
\left\{ \psi _{1}(x),\psi _{2}^{*}(y)\right\} &\rightarrow &\pm m\,\,\,\, 
\nonumber
\end{eqnarray}
where the $x,y$ in the last relation are on the upper/lower horizon so that
the distance is $\pm $time-like. In this case no algebraic reprocessing is
needed, we obtain two conformal theories with different chirality which
applied to the vacuum generate $H_{\pm }$ subspaces.. The fact that they did
not simply originate from a mass zero limit shows up in their nontrivial
m-dependent relative commutation relation which is the memory of the
relative time-like difference. In fact the light cone restriction is a
peculiar chiral conformal theory, a kind of hidden massive theory, which by
the adjunction of the opposite light cone translation can be cropped into a
wedge-localized massive theory 
\begin{equation}
alg\left\{ \mathcal{M}_{+},U_{-}(a\geq 0)\right\} =\mathcal{M}
\end{equation}
In fact these properties of simple models are believed to be a general
structural features of light ray restrictions, a point to which we plan to
return in a separate paper.

As mentioned before, all modular constructions for factorizing theories can
be based on the algebraic structure as well as the vacuum annihilation
properties of the Z-operators. After having constructed the local net one
would like to check however the interpretation of the n-particle Z-states in
terms of incoming states by establishing an operator formula (which
generalizes the well-known exponential formula for diagonal S-matrices)
based on the (Haag-Ruelle, LSZ) time dependent scattering theory. The
explicit knowledge of the local double cone net together with the isolated
one particle states (the spectral gap property) allows for the construction
of the incoming fields in terms of the $Z^{\prime }s$ and by inversion also
the (a priori unknown for the nondiagonal models) formulas for the $%
Z^{\prime }s$ in terms of the incoming free fields. At this point the
correctness of the scattering interpretation of the coefficient functions (%
\ref{series}) in terms of formfactors between scattering states is confirmed
by selfconsistency.

\section{Outlook}

We have succeeded to reformulate the existence of ``quantum localized''
PFG's in interacting QFT in terms of crossing symmetric on-shell operators $%
M(\theta )$ which appear in the commutation relation between the $Z^{\prime
}s$ and their Tomita transformed $JZJ$ and we identified these operators
with well-defined (through their commutation relations) expressions $%
T(\theta )$ within the more restricted context of integrable models. As a
side result we obtained a relation between the extremely deep crossing
symmetry\footnote{%
In fact it was the ill-understood (outside some formal perturbative
calculations) crossing symmetry, a kind of TCP relation which is not global
but holds for individual particles, for which the desire for a more profound
understanding led first to Veneziano's dual model and then (after the
original motivation was forgotten) to string theory.}. It would have been
somewhat disappointing if the understanding of this most characteristic and
profound aspect of QFT could have been achieved by already well-known
properties as Haag-Ruelle scattering theory and the thermal KMS
condition.without the necessity to introduce a new concept as the role of
PFW's and their connection with an associated modular light ray theory.

Two questions immediately arise: can one use this gain of conceptual insight
for the improvement of the Bootstrap-Formfactor program, and secondly could
the new algebraic structure based on the general possible existence of PFG's
be useful for a kind of ``modular revival'' (this time build on
noncommutative Tomita modular theory instead of the commutative measure
theory underlying euclidean functional integrals) of the old constructive
program to get beyond free fields and perturbation theory?

As far as the first question is concerned, we remark that the outstanding
problem is the physical understanding and the mathematical description of
the ``operator content'' of local fields. By this we mean the understanding
of their expansion coefficients in terms of an auxiliary basis (\ref{series}%
) which leads to formfactors. As it turns out, this is a very difficult task
even for the simplest case of (diagonal Toda) factorizing field theories.
The identification of the sequences of symmetric polynomials which
characterize the formfactors of the list of (composite) fields up to this
date has not been accomplished. Our algebraic viewpoint which shifts the
emphasis from individual field coordinates to double cone algebras, reveals
that the traditional field construction carries a lot of nonintrinsic
structure which adds little to the securing of the existence and the
understanding of the important dynamical properties of these models. The
construction of a useful basis of local fields (i.e. a basis in the local
field space), which parallels the construction of Wick polynomials in the
free case, is a highly technical problem with a large amount of
arbitrariness. This is of course a phenomenon which had to be expected on
the analogy with coordinates in differential geometry. In fact in
(nonperturbative) LQP the technical difficulties in using field coordinates
versus intrinsically defined local nets if anything become even more awkward
than the use of coordinates in differential geometry. Perturbation theory
which requires the selection of a set of field coordinates is no good guide
for these aspects, but even there one realizes the clumsiness of field
coordinates if one tries to understand the insensitivity of the S-matrix (in
the present setting its role as a relative modular invariant for wedge
algebras) against local changes of field coordinates. Even formulations
which are usually considered to be the quintessence of the nonperturbative
approach as e.g. the Wightman approach are not completely intrinsic. In the
present work we have used modular ideas which force us to understand the
existence and physical content of QFT in terms of net of von Neumann
algebras. The crucial step, the construction of the double cone algebras, is
facilitated by the use of (far off-shell) light ray algebras which turned
out to be ``hidden'' chiral conformal theories. This approach, whose
complexity increases with the distance from on-shell properties (or distacne
to polarization free objects), is particularly suited for high energy
particle physics where the main measurable quantities are not correlation
functions of fields but rather the S-matrix and perhaps formfactors of some
distinguished fields (Noether currents) both closely related to modular
concepts associated with wedge localization. The situation is different in
statical mechanics and condensed matter physics; in that case one has to
work harder for the calculation of correlation functions (which often can be
measured by neutron scattering). Our nonperturbative approach contains the
message to separate the construction of a nontrivial field theory and the
understanding of its physical content from the technically complicated
construction of its field coordinates.

A somewhat related problem is the question of how to avoid an explicit
formula for the PFG operators in terms of incoming operators. Such a formula
is a priory not known (i.e. cannot be written down on the basis of
commutation relations which involve the S-matrix data) for nondiagonal
factorizing models. A glance at the $Z^{\#}(\theta )$ operators reveals that
if the annihilators annihilate the vacuum state, the Z-F algebra determines
n-particle Hilbert space inner products (even without knowing a relation to
incoming fields). For the interpretation in terms of formfactors it is
sufficient to be able to relate the vector \textit{states} generated by the $%
Z^{\#}(\theta )$ to the incoming/outgoing states. The construction of the
double cone algebras can be done in the related total space without use of a
Fock space structure because the expansion formulas (\ref{series}) are
defined as soon as the iterative action of the $Z^{\#}(\theta )^{\prime }s\,$%
on the vacuum is known. Having the local algebras within that space, one
only has to remind oneself that the Haag-Ruelle scattering theory uses just
locality and the energy-momentum spectrum. Therefore one may \textit{compute
the incoming fields} and then invert the relation in favor of writing the
Z's in terms of $a_{in}^{\#}.$

The second question concerning the applicability of these ideas outside of
factorizing models should be subdivided according to space-time dimension $%
=1+1$, or $>1+1$. In the first case the mass shell is still parametrized in
terms of rapidities but they do not furnish a uniformization variable for
the S-matrix which retains its rich creation/annihilation threshold analytic
structure. The prerequisites for quantum wedge locality hold since crossing
symmetry remains still a valid concept. Of course this does not yet give a
blueprint for the construction of $Z$ and $M(\theta ).$ As in the
factorizing case this cannot be fully covariant; at least the pointlike
parity transformation must be violated. This reduced covariance of the PFG
operators is related to their weaker localization and not with broken
symmetries in the final QFT).

As a curious side remark we mention that the central idea of this paper,
namely the existence of PFG operators as the best compromise between the
of-shell localization of fields with the on-shell particle structure, has
some vague resemblance with certain formal simplifications in the so-called
light-cone or light-front quantization. Closing one's eye to the fact that
the canonical nature of the approach (as in all quantization approaches) is
not compatible with realistic ideas about interactions\footnote{%
In fact genuine 2-dim. conformal fields, which are sums over products of
anomalous dimension (spin) chiral fields from the two light cones, already
expose this short-distance problem in the light ray restriction of pointlike
fields.} (apart from superrenormalizable interactions), one realizes that
formally such a quantization suppresses vacuum or one-particle polarization
phenomena \cite{Heinz} which is of course precisely the characteristic
property of the PFG's. However whereas in the light-front approach the
relation of the so obtained field operators to the original local fields
seems to have been irrevocably lost, and with it the possibility if
physically interpreting the theory (apart from the one particle spectral
informations), the PFG'ss are auxiliary operators which have a conceptually
clear and mathematically precise relation within LQP. The message from our
modular approach is that via ``quantum localization'' (formation of algebras
and intersections) one should be able to recover the local variables from
the ones obtained by light cone quantization. For people who do not believe
in strange coincidences we may add that the Galilei group which appeared as
a Poicar\'{e} subgroup in light cone quantization \cite{Heinz} had an
independent outing in form of an 8-parametric subgroup in the modular
approach in connection with so called modular intersections (see the first
paper in \cite{papers}).

As (apart from superrenormalizable theories, where the fields are ``locally
normal'' with repect to the canonical free fields) the only consistent
remainder of the standard canonical quantization formalism is the Einstein
causality, the semilocal PFG' fields of this paper could have a similar
relation to the light cone formalism fields. In view of the popularity which
ideas of noncommutative geometry have enjoyed within part of the physics
community, the present message, that the nonperturbative construction of
local fields requires very noncommutative intermediate steps based on
modular theory, should have an attractive appeal.

The higher dimensional cases are additionally complicated by the fact that
wedge-adapted rapidities do not furnish a complete mass shell
parametrization and that even if one succeeds to construct PFG's, the
compactly localized double cone algebras cannot be obtained by intersecting
only two wedge algebras with the translated opposite (i.e. J-transformed)
wedge algebras. Here a modular revival of the idea inherent in the old dual
model, namely a relation between a higher dimensional (on-shell) S-matrix
and a low dimensional d=1+1 (off-shell) field theory would be desirable.

Acknowledgements: The authors thank M. Karowski for explaining some some of
the more intricate features of the formfactor program and K.-H. Rehren for
critical reading of an early draft version.


\begin{thebibliography}{99}
\bibitem{Schro1}  B. Schroer, Nucl. Phys. \textbf{B 499}, (1997) 519, 547

\bibitem{Schro2}  B. Schroer, ``Modular Wedge Localization and the d=1+1
Formfactor Program'', hep-th/9712124, to be published in AOP

\bibitem{Bru-Gui-Lo}  private discussion with D. Guido. We also refer to a
recent unpublished manuscript by R. Brunetti, D. Guido,R. Longo, ``On the
Intrinsic Construction of Free Theories via Tomita Takesaki Theory''.

\bibitem{papers}  B. Schroer and H.-W. Wiesbrock, ''Modular Theory and
Geometry'' math-ph/9809003; B. Schroer and H.-W. Wiesbrock, ``Looking beyond
the Thermal Horizon: Hidden Symmetries in Chiral Models'', in preparation

D. Buchholz, O. Dreyer, M. Florig, S. J. Summers, ``Geometric Modular Action
and Spacetime Symmetry Groups'', math-ph/9805026

\bibitem{Nieder}  Max Niedermaier, Nucl. Phys. B519 (1998) 517,
hep-th/9706172 to be published in CMP.

\bibitem{Bisognano-Wich}  J. Bisognano and E. H. Wichmann, J. Math. Phys. 
\textbf{17}, (1976) 303

\bibitem{Bog}  N. N. Bogoliubov, A. A. Logunov, A. I. Oksak and I. T.
Todorov, ``General Principles of Quantum Field Theory'' Kluwer Academic
Publishers 1990, anf references therein.

\bibitem{Wein}  S. Weinberg, ``The Quantum Theory of Fields, I,
Foundations'', Cambridge University Press, 1955

\bibitem{E-G}  H. Epstein and V. Glaser, Ann. Inst. H. Poincar\'{e} Sect. 
\textbf{A 19}, (1973) 211

\bibitem{Kou-Mu}  A. Koubek and G. Mussardo, Phys.Rev.Lett. \textbf{B316},
(1993) 193\thinspace

A. Fring, G. Mussardo and P. Simonetti, Nucl. Phys. \textbf{B393}, (1993) 413

M. Pillin, ``The Formfactors in the Sinh-Gordon Model'' hep-th/9712033

\bibitem{Haag}  R. Haag, ``Local Quantum Physics'', Springer Verlag 1992

\bibitem{Mueger}  M. Mueger, Commun. Math. Phys. \textbf{191}, (1998) 137

\bibitem{Karowski}  H. M. Babujian, A. Fring, M. Karowski and A. Zappletal,
``Form Factors in Integrable Quantum Field Theories: the Sine-Gordon
Model'', hep-th/9805185

\bibitem{Smirnov}  F. A. Smirnov, Adv.Series in Math. Phys. \textbf{14},
(World Scientific, Singapore, 1992)

\bibitem{Zam}  A. B. Zamolodchikov and Al.. B. Zamolodchikov, Ann. Phys. 
\textbf{120}, (1979) 253

L. D. Faddeev, Sov. Sci, Rev. Math, Phys. \textbf{C1, }(1980) 107

\bibitem{Lash}  M. Yu Lashkevich, ``Sectors of Mutually Local Fields in
Integrable Models of Quantum Field Theory'', hep-th/9406118

\bibitem{St-Wi}  R. F. Streater and A. S. Wightman, ``Spin, Statistics and
all That''. Benjamin Inc. New York 1964

\bibitem{Bo-Zi}  H.- J. Borchers and W. Zimmermann, Nuovo Cimento \textbf{31}%
, (1963) 1047

\bibitem{Kar}  M. Karowski, H. J. Thun, T. T. Truong and P. Weisz, Phys.
Lett. \textbf{B67}, (1977) 321

\bibitem{Wies}  H.-W. Wiesbrock, Lett. Math. Phys. \textbf{184}, (1997) 683

\bibitem{Sew}  G. Sewell, Ann. of Phys. \textbf{141,} (1982) 201

S. Summers and R. Verch, Lett. Math. Phys. 37, (1996) 145

\bibitem{Guido}  D. Guido, R. Longo and H.-W. Wiesbrock, Commun. Math. Phys. 
\textbf{192}, (1998) 217

\bibitem{Heinz}  T. Heinzl, ``Light-Cone Dynamics of Particles and Fields'',
chapter4, hep-th/9812190
\end{thebibliography}
\end{document}